\documentclass[twocolumn,showkeys,preprintnumbers,floatfix,amsmath,a4paper,amssymb,nofootinbib]{revtex4}

\usepackage{amsmath}
\usepackage{graphicx}
\usepackage{dcolumn}
\usepackage{bm}
\usepackage{amssymb}
\usepackage{epstopdf}

\begin{document}
\def\P{\mathbf{P}}
\def\Q{\mathbf{Q}}
\newcommand{\ket}[1]{\left| #1 \right>} 

\title{Chromatic transitions in the emergence of syntax networks}

\author{Bernat Corominas-Murtra$^{1,2,3}$, Mart\'i S\`anchez Fibla$^4$, Sergi Valverde$^{2,3}$ and Ricard Sol\'e$^{2,3,5}$
}

\affiliation{
$^1$ (1) Institute of Science and Technology Austria, Am Campus 1, A-3400, Klosterneuburg, Austria\\
$^2$Complex Systems  Lab, ICREA-Universitat Pompeu Fabra,   Dr    Aiguader   88,   08003   Barcelona,   Spain\\
$^3$Evolution of Technology Lab, Institut de Biologia Evolutiva (CSIC-UPF), Passeig Maritim de la Barceloneta, 37-49, 08003 Barcelona, Spain\\
$^4$SPECS, Technology Department, Universitat Pompeu Fabra, Roc Boronat 138, 08018 Barcelona, Spain\\
$^5$ Santa Fe Institute; 1399 Hyde Park Road; Santa Fe; NM 87501; USA
} 


\begin{abstract}
The emergence of syntax during childhood is a remarkable example of how complex correlations 
unfold in nonlinear ways through development. In particular, rapid transitions seem to occur as 
children reach the age of two, which seems to separate a two-word, tree-like network of syntactic 
relations among words from a scale-free graphs associated to the adult, complex grammar. Here we explore
the evolution of syntax networks through language acquisition using the {\em chromatic number}, which captures the 
transition and provides a natural link to standard theories on syntactic structures. The data analysis 
is compared to a null model of network growth dynamics which is shown to display nontrivial and sensible 
differences. In a more general level, we observe that the  chromatic classes define independent regions of
the  graph, and thus,  can be  interpreted as  the footprints  of incompatibility relations, somewhat as opposed 
to modularity considerations. 
\end{abstract}
\maketitle


\section{Introduction} 
\label{Introduction}


The origins of human language have been a matter of intense debate. 
Language is a milestone in our evolution as a dominant species and is likely 
to pervade the emergence of cooperation and symbolic 
reasoning  \cite{Maynard:1995, Bickerton:1990, Hauser:2002, Christiansen:2003}. 
Maybe the most defining and defeating 
trait is its virtually infinite generative potential: words and sentences can be constructed 
in recursive ways to generate nested structures of arbitrary length \cite{Chomsky:1988, Hauser:2002}. Such structures 
are the product of a set of rules defining syntax, which are extracted by human brains 
through language acquisition during childhood after a small sample of the whole combinatorial universe of sentences 
has been learned. And yet, in spite of its complexity, syntax is accurately acquired by children, 
who master their mother tongue in a few years of learning. Indeed, around the  age of two, linguistic structures produced
by  children   display  a  qualitative  shift   on  their  complexity,
indicating a  deep change on  the rules underlying them \cite{Radford:1990, Corominas-Murtra:2009a}.   
This sudden increase of grammar complexity is  known as the {\em syntactic spurt},
and  defines the edge between  the {\em  two words}  stage,  where only
isolated words  or combinations of two words occur, to  a stage where
the grammar  rules governing this syntax  are close to the  one we can
find in adult speech -although  the cognitive maturation of kids makes
the  semantic content or  the pronunciation  different from  the adult
one. How can we explain or interpret such nonlinear pattern?
\begin{figure*}[ht!]
\includegraphics[width=17.2cm]{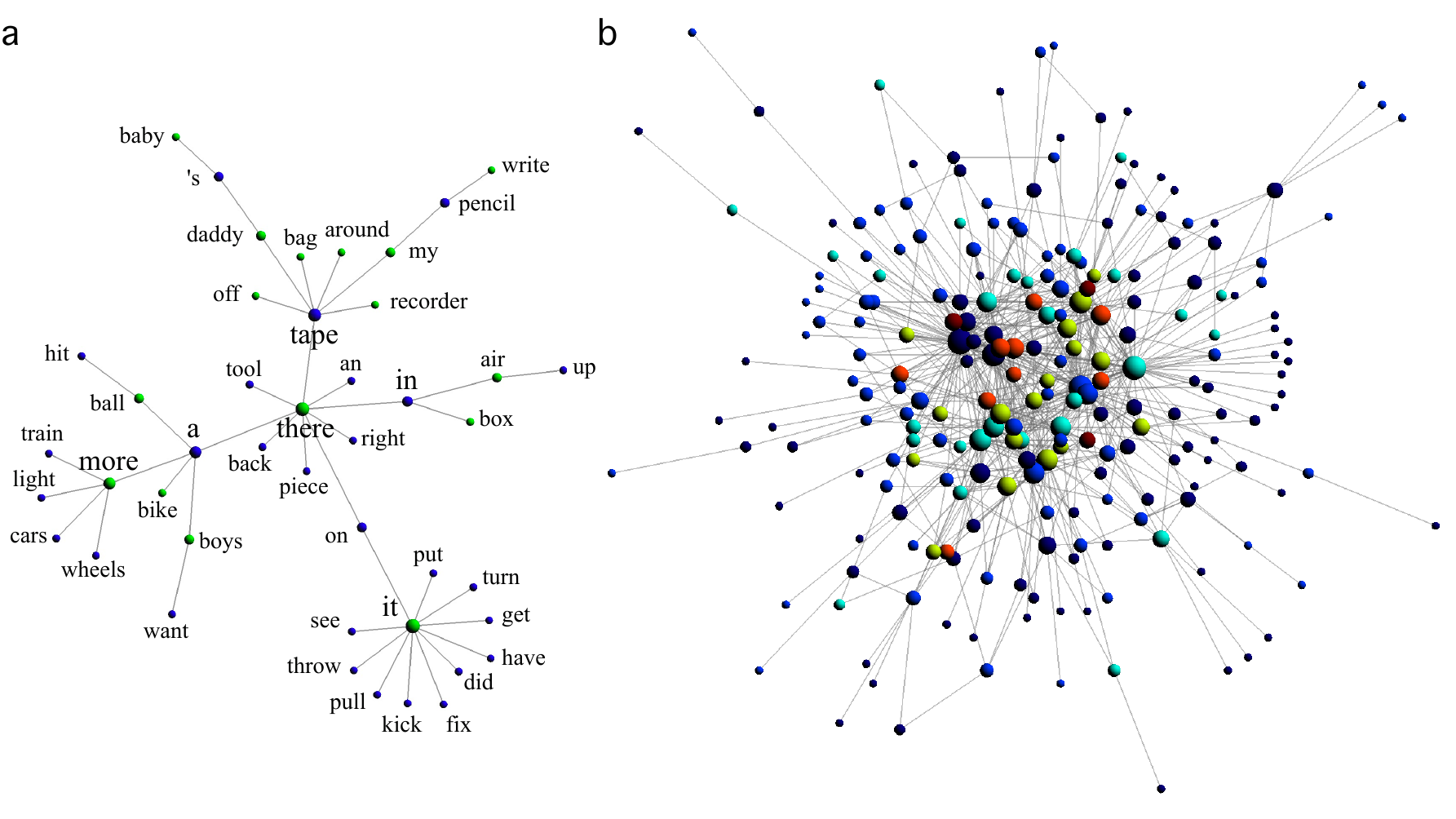}
 \caption{Optimal colourings of syntactic networks before and after the syntactic spur. 
(a) A syntactic network before the transition (3th corpus) is largely bipartite (this network accepts a
 2-coloring).  (b) Post-transition network (7th corpus) is remarkably more complex, which 
 corresponds to high chromatic number $\chi({\cal G}_{7})=6$.  All networks coming from
  Peter dataset. Time spent between these two corpora is about two months and  a half (see text).}
\label{Fig:Colors}	
\end{figure*}



Statistical physicists have approached the problem of language evolution showing for example that nontrivial patterns are 
shared between language inventories (collections of words) and some genetic and 
ecological neutral models \cite{Blythe:2007} --see \cite{Sole:2010b} and references therein. However, most of these models 
do not make any assumption about the role played by actual interactions among 
words, or, more generally, linguistic units,  which largely define the nature of linguistic structures.  In this context, a promising approach to its structure and evolution involves considering language in terms of networks of interconnected units instead of unstructured collections of elements (e.g., words or syllabes) \cite{Sole:2010}. In this context, syntactic networks, in which nodes are words and links the projection of actual syntactic relations, have been shown to be an interesting abstraction to grasp general patterns of language production \cite{Ferrer:2004, Steele:2009, Sole:2010}. Specially valuable has been the quantitative data obtained from syntax networks obtained along the process of syntax acquisition \cite{Corominas-Murtra:2009a, Ke:2008, Barcelo-Coblijn:2012}.

At the fundamental level, syntax can be understood as a set of symbols associated under 
 a restricted universe of combinations somewhat similar to chemistry. Atoms and words 
 would then be linked through compatibility relations defining what can be 
 combined and what is forbidden. The power of this picture is supported by the use of linguistic methods in 
 the systematic characterization of chemical structures \cite{Rankin:1971}. 
 Chemical structure diagrams can thus be seen as some sort of language, with 
 chemical species and bonds as key ingredients. In a more abstract fashion, 
 we can say that general rules of combining elements within a given set of interacting 
 pieces with well defined functional meaning is at work in both language and chemistry. 
 
Following the chemical analogy, where abstract classes of "nodes" can be 
defined, we will take advantage of graph colorability theory as a general 
framework to detect transitions based on qualitative changes of compatibilities.  
Specifically, we  suggest that a new combinatorial approach grounded on a graph 
colouring may enable a better  understanding of the evolution of 
networks having internal relations of compatibility 
 (e.g., some kind of syntactic rules).  In this context, we  propose  the {\em chromatic number} 
 -and associated measures- of the graph \cite{Tutte:1941, Bollobas:1998,Bollobas:2001} 
 as an indicator of network complexity. Within our context, it will be a surrogate 
 of syntactic complexity. The chromatic number is defined as the minimal number of {\em colors} 
needed to {\em paint} all nodes of the graph in a way that no adjacent nodes
 have the same color \cite{Bollobas:1998}.  The $q$-coloring problem, i.e., to 
 know whether a graph can be colored with $q$ different colors is one of the
  most important $NP$-complete problems.  From the statistical physics point of
  view, an analogous problem is defined within the context of the {\em Potts} model \cite{Wu:1982}. 
  
It is worth to emphasize that transitions  in the evolution of the chromatic number have been widely studied
 in models of random graphs  \cite{Bollobas:1988, Bollobas:2001, Achiloptas:1999, Zdeborova:2007}.
 Our exploration over sequences of syntactic graphs mapping child language
acquisition also displayed transitions in the chromatic number (see below). This is, to 
the best of our knowledge, the first time that such transitions have been reported 
in a real system. Classes of nodes would be defined precisely by the fact that there
 are no connections among them, a measure conceptually opposite to graph modularity.
   
The remaining of the paper is organized as follows. Section \ref{Colouring} is devoted 
to a brief revision of the so-called {\em Potts model} as the way to introduce the chromatic number. 
In section \ref{Sec:Evolution} we apply these theoretical constructs to our problem 
and we analyze the obtained data by using different estimators of relevance, the
 most prominent of them being a null model of random sentence generation. 
 In section \ref{Discussion} we discuss the obtained results and we highlight a number
 of potential impacts of this kind of complexity estimators for complex networks.
  


\section{Graphs and Coloring: Basics}
\label{Colouring}


%

We will work over undirected graphs. An undirected graph  ${\cal G}(V, E)$ -hereafter,  ${\cal G}$- is  composed by the
set  of $V=\{v_1,...,v_n\}$ {\em  nodes} and  a set
$E= \{ e_j \| 1 \le j \le m \} \subseteq V\times V$ of edges.  Each (unordered) pair $e_j=\{v_i,v_k\}$ 
depicts a link between nodes  $v_i$ and $v_j$.  The number of links $k(v_i)$ 
  attaching node $v_i$ is the  {\em degree} of the  
node and $\langle  k\rangle$ is the {\em average degree} of the graph ${\cal G}$.  
The {\em degree distribution}  $P(k)$ accounts for the probability to select a node 
at random having degree $k$.  The identity  card of  a graph  is the  
 so-called {\em  Adjacency matrix}, $\mathbf{a}({\cal G})$, which is defined as follows:
 
\begin{equation}
a_{ij}=\left\{\begin{array}{ll}
1,\;{\rm iff}\;(\exists e_k\in E):(e_k=\{v_i, v_j\})\nonumber\\
0,\;{\rm Otherwise}\quad.\nonumber
\end{array}
\right.
\end{equation}
We observe that the adjacency matrix of undirected graphs is symmetrical, i.e.,  $a_{ij} = a_{ji}$.




The computation of the chromatic number can be formulated as the following
combinatorial problem:  What is the minimal number of 'colours' needed 
to paint all nodes of  the graph in such a way that no single node is connected to
 neighbors having the same color? We can map this problem into the antiferromagnetic
$q$-dimensional Potts model at  $T=0$ \cite{Wu:1982}.   This model is a 
generalization of the classical Ising model for lattices: at every node of this lattice we
 place a particle having a spin which energetically constraints the state  of  its  neighbors.  
Traditionally, spins can have  only  two states,  namely $\ket{\uparrow}$ and $\ket{\downarrow}$. 
In the Potts model,  compatibility relations  take into account  an arbitrary number $q >2$ of
 different states. 
Let us consider a partition of nodes $V$ containing  $q$
different classes, namely, $G_q(V)=\{g_1, ..., g_q\}$ of $V$, i.e.:
\begin{equation}
\bigcap G_q=\varnothing\;{\rm and}\;\bigcup G_q=V\quad,
\end{equation}
The  {\em state}  $\sigma_i$ of  node $v_i$ indicates the class  of
$G_q(V)$ to which the node belongs to, i.e., $\sigma_i\in g_j$. 
 Let ${\cal F}_q(V)$ be  the ensemble  of all partitions  of $V$ 
 containing $q$ different  classes. Every element in ${\cal  F}_q(V)$  
 has the following energy 
 penalty\footnote{In our approach, the energy units of this Hamiltionian are arbitrary.}:
\begin{equation}
{\cal H}(G_q)=J\sum_{i<j}a_{ij}\delta(\sigma_i, \sigma_j)\quad,
\label{Hamilton}
\end{equation}
where $J=1$ is  the {\em coupling constant} and $\delta$ is the 
Kronecker symbol:
\begin{equation}
\delta(\sigma_i,\sigma_j)=\left\{\begin{array}{ll}
1,\;{\rm iff}\;i=j\nonumber\\
0,\;{\rm Otherwise}\quad.\nonumber
\end{array}
\right.
\end{equation} 

Intuitively, the higher  the presence of pairs of connected nodes belonging to  the 
same  state, the  higher will be  the energy  of the global  state of  
the graph.  Given  a fixed  $q$, the  configurations displaying  minimal  
energy  may  have  an   amount  of  non-solvable situations, leading to
 the  unavoidable presence of connected nodes at the same state.  
 This phenomenon  is called {\em frustration}, and for these configurations,  
 the ground state of the  Hamiltonian defined in (\ref{Hamilton}) displays  
 positive energy.  If there is no frustration,  i.e., $\exists G_q\in {\cal F}_q(V)$,
  we can find a partition that satisfies: 
\begin{equation}
{\cal H}(G_q)=0\quad,
\end{equation}
and we say  that the  graph is $q$-colorable,  being the $q$ different {\em  colors} 
the $q$ different classes or members of $G_q$. When the graph is $q$-colorable, 
there is at least one partition $G_q\in{\cal F}_q(V)$ such that, if $v_i, v_j\in  V$ 
belong  to  the same  {\em  class} or  {\em  color} of  the partition, 
namely $g_l\in G_q$. We deduce that: 
\begin{equation}
(v_i, v_j\in g_l)\Rightarrow a_{ij}=0\quad.
\label{Nocontact} 
\end{equation}
Relation  (\ref{Nocontact}) maps color classes onto disjoint sets of graph elements 
(adjacent nodes have a different color). Now, the coloring problem consists  in finding 
the minimal number of classes (or colors) required to properly {\em paint} the graph. 
This is the so-called {\em Chromatic Number}  of the graph ${\cal G}$:
\begin{equation}
\chi({\cal G})=\min\{q:(\exists G_q\in {\cal F}_q(V)):{\cal H}(G_q)=0\}\quad.
\label{HamiltonColor}
\end{equation}
Now suppose network  partition(s)
  $G^*_q\in {\cal F}_q(V)$ having minimal energy, see equation (\ref{Hamilton}), given a number of colours $q$:
\begin{equation}
G^*_q=\min_{G_q\in {\cal F}_q(V)}\{{\cal H}(G_q)\}\quad.
\end{equation}
In general, the process of search for the chromatic number yields a decreasing sequence of energies ending  at ${\cal H}(G^*_{\chi({\cal G})})=0$:
\begin{equation}
{\cal H}(G^*_1)\leq ...\leq {\cal H}(G^*_{\chi({\cal G})})=0\quad,
\label{sequence}
\end{equation}
In order to assess the statistical significance of chromatic numbers, 
 we define the  {\em relative energy} of any $q$-coloring as follows:
\begin{equation}
f_{q}(\chi)=\frac{ {\cal H}(G^*_{q})}{|E|}\quad,
\label{fchi} 
\end{equation}
where $|E|$ is the number of edges in the graph $G$. This quantity 
$0 \le f_{q}(\chi) \le 1$  corresponds to the minimal (relative) number 
of frustrated links or {\em violations} (i.e., when adjacent nodes have the same color). 

\begin{table*}
\begin{tabular}{c|ccccccccccc}\hline
                    &${\cal G}_{P1}$&${\cal G}_{P2}$ &${\cal G}_{P3}$ &${\cal G}_{P4} $&${\cal G}_{P5}$ &${\cal G}_{P6}$ &${\cal G}_{P7}$ &${\cal G}_{P8}$ &${\cal G}_{P9}$ &${\cal G}_{P10}$ &${\cal G}_{P11}$ \\\hline
$f_1(\chi)$ &  $1$		 &  $1$	         &  $1$            & $1$               & $1$              & $1$                  &       $1$               &            $1$             &   $1$                     &   $1$           & $1$            \\ 
$f_2(\chi)$ &  $0$               &   $0$             &  $0$             &  $1/49$        & $ 5/105$     & $66/434$        &     $131/644$    &          $87/589$     &   $157/903$        &  $104/659$&$95/717 $  \\
 $f_3(\chi)$ & $0$               &   $0$             &  $0$              & $0$              &  $ 0$             & $8/434$          &     $31/644 $     &          $15/589$     &   $ 40/903$         &  $20/659$   & $10/717$  \\
$f_4(\chi)$ & $0$               &   $0$             &  $0$               & $0$              &  $ 0$             & $0$                   &     $8/644$       &              $0$            &   $ 8/903$            &  $ 2/659$     & $0$            \\
$f_5(\chi)$&  $0$               &   $0$             &  $0$               & $0$              &  $ 0$             & $0$                   &     $1/644$      &               $0$            &   $0$                    & $0$               & $0$  \\
\hline
                    &${\cal G}_{C1}$&${\cal G}_{C2}$ &${\cal G}_{C3}$ &${\cal G}_{C4} $&${\cal G}_{C5}$ &${\cal G}_{C6}$ &${\cal G}_{C7}$ &${\cal G}_{C8}$ &${\cal G}_{C9}$ &${\cal G}_{C10}$ &${\cal G}_{C11}$ \\\hline
$f_1(\chi)$ &  $1$		 &  $1$	         &  $1$            & $1$               & $1$              & $1$                  &       $1$               &            $1$             &   $1$                     &   $1$           & $1$            \\ 
$f_2(\chi)$ &  $6/140$          &   $5/119$         &  $11/156$   &  $6/128$      & $10/152$   & $14/199$        &     $61/361$    &          $65/442$       &   $71/439$           &  $93/592$  &$131/687$  \\
 $f_3(\chi)$ & $0$               &   $0$             &  $0$              & $0$              &  $ 0$             & $0$                  &      $9/361 $     &          $11/442$     &   $ 8/439$         &  $16/592$   & $29/687$  \\
$f_4(\chi)$ & $0$               &   $0$             &  $0$               & $0$              &  $ 0$             & $0$                   &     $0$            &              $0$               &   $ 0$                   &  $1/592$     & $4/687$         \\  
\hline
\end{tabular}
\caption{Relative energy values of $q$-colorings in the Peter (top) and Carl (bottom) datasets (see text). }
\label{TaulaColoring}
\end{table*}

Despite  the high  complexity of this problem (computing the  chromatic number in an 
arbitrary  graph is a $NP$-hard problem) several bounds can be defined. A lower bound can be defined from the so-called {\em Clique number}. 
A {\em clique} is a 
subgraph in which every node is connected to all other nodes in the subgraph. 
The {\em Clique number}  $\omega({\cal G})$ is  the size of  the largest clique 
in the graph, which is a natural lower bound for $\chi({\cal G})$ \cite{Bollobas:1998}:
\begin{equation}
\omega({\cal G})\leq \chi({\cal G})\quad.
\label{clique}
\end{equation}
Alternatively, an  upper bound  on $\chi({\cal G})$  can be  defined by
looking at the $K$-core structure  of ${\cal G}$. The $K({\cal G})$ 
core is the largest subgraph whose nodes display degree higher or equal
to $K$. Now, lets $K^*({\cal G})$ be the $K$-core  with largest
 connectivity that  can be  found in ${\cal G}$: 
\begin{equation}
K^*=\max\{K:K({\cal G})\neq \varnothing\}\quad.
\label{Kcore}
\end{equation}
Then, it can be shown that $K^*$ sets an
 upper bound to the chromatic number \cite{Bollobas:1998}:
\begin{equation}
\chi({\cal G})\leq K^*+1\quad.
\label{kcore}
\end{equation}
Finally, let us mention that, for some families of random graphs the chromatic number has  an asymptotic  behavior depending on the
average connectivity \cite{Bollobas:2001},
 $\chi({\cal G})\sim\frac{\langle k\rangle}{\log \langle k \rangle}$ 
However, the above relationship does not hold for scale-free networks 
with exponent $2<\gamma<3$. These heterogenous networks cannot have a
 stable value of the chromatic number because their clique number (\ref{clique}) 
 diverges with the graph size, even at constant $\langle k\rangle$ \cite{Biancomi:2006}.

\section{The Evolution of $\chi$ along syntax acquisition}

\label{Sec:Evolution}


Here we study the evolution of the chromatic number through language 
development as captured by syntax graphs. 
 We compare the chromatic number with the lower and upper bounds provided by the
clique number and the maximal $K$-core, respectively.  We assess the 
relevance of computed chromatic numbers with the corresponding minimal energy. The 
combination of these two measurements enable us to interpret the nature of the chromatic 
number. Specifically,  we can check wether changes in this number reflects a global pattern or instead 
some anomalous  behaviour of a small, localized subgraph. Finally, we provide
further validation of our analysis by  comparing chromatic numbers in empirical and
synthetic networks obtained through a random sentence generator.
  
  \subsection{Building the Networks of Early Syntax}
\label{Construction}

Through the process, networks built upon the aggregation of syntactic structures from child's productions grow and change 
in a smooth fashion until a rapid transition occurs  \cite{Corominas-Murtra:2009a, Bernat:2007, Barcelo-Coblijn:2012} 
--see also \cite{Sole:2010}.  We reconstruct syntactic networks  by projecting the raw constituent structure, i.e., 
 phrase structure of children's utterances, into linear relations among lexical items
 \cite{Popescu:2003}.  Then, we aggregate all these
 productions in a single graph where nodes are lexical items and links 
 represent syntactic relations between them
 \cite{Corominas-Murtra:2009a, Bernat:2007}. These networks provide a unique
  window  into the patterns of change occurring in the language acquisition process.

The two cases studied here are obtained from the CHILDES Database 
\cite{Bloom:1974, Bloom:1975} which includes conversations between children 
and parents. Specifically, we choose Peter and Carl's corpora, whose 
structure has been accurately extracted and curated. 
For both Peter and Carl's corpora, we choose $11$ different recorded conversations 
distributed in approximately uniform time intervals ranging from the age of $\sim 20$ months 
to the age of $\sim 28$ months. The chosen interval corresponds to the period in which the
 syntactic spurt  takes place.  From every recorded conversation, we extract the
syntactic network of child's utterances obtaining a sequence of $11$ syntactic graphs 
corresponding  to the sequence of Peter conversations ${\cal G}_{P1},..., {\cal G}_{P11}$
 and Carl's conversations ${\cal G}_{C1},..., {\cal G}_{C11}$.


\subsection{Chromatic transition from bipartite to multicoloured networks} 
\label{Real}
 \begin{figure*}
\begin{center}	%
\includegraphics[width=16.5cm]{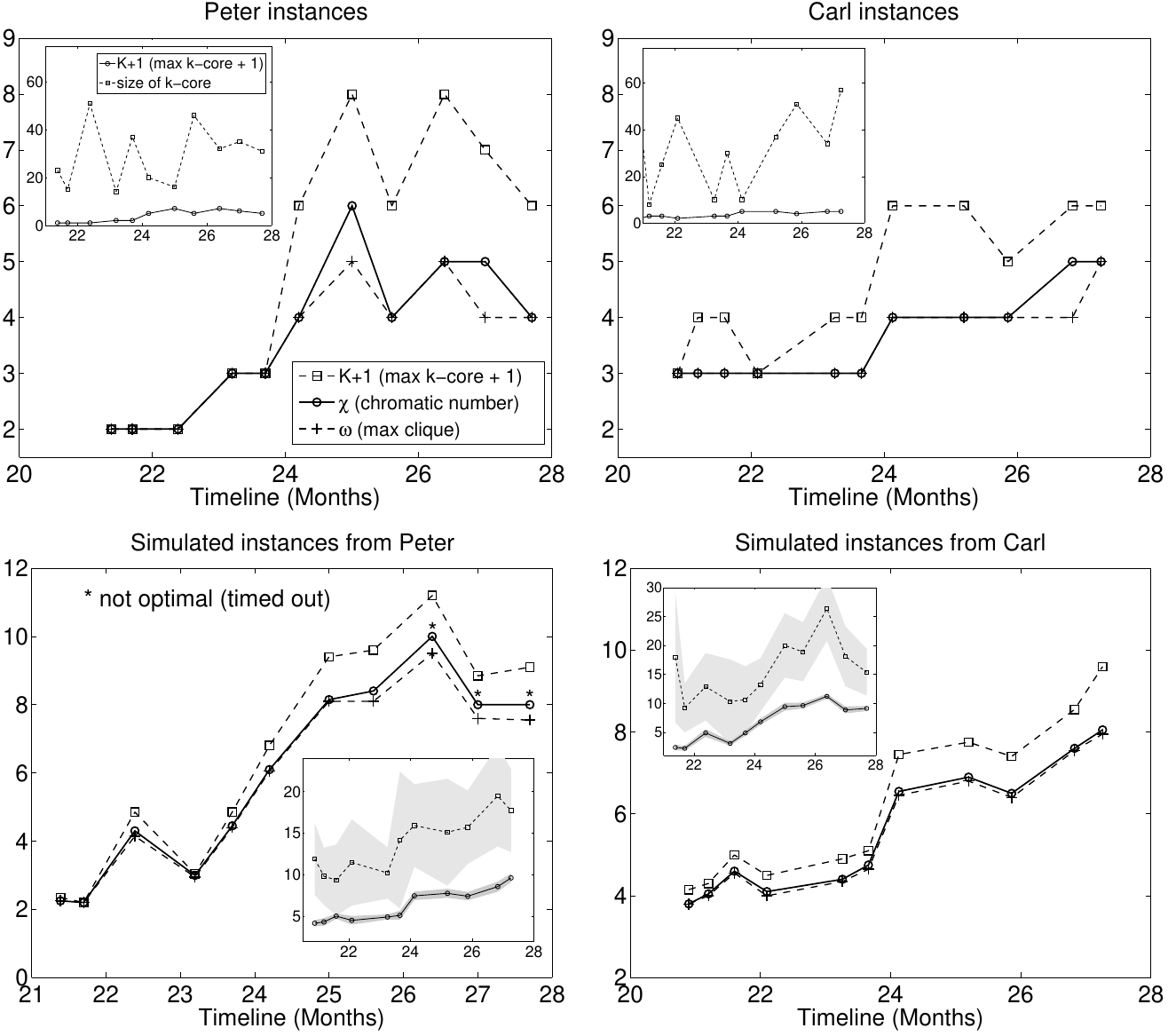}
\caption{ Evolution of the chromatic number $\chi$ (solid line), maximum clique $\omega$ 
(dashed line with crosses) and the maximum core $K^* +1$ (dashed line with squares)
in (top) real networks and in (bottom) an ensemble of $n=20$ simulated networks (see text). 
Inset: Comparison between the time evolution of the $K$-core size (dashed line) and the
maximum $K^*$-core size (solid line). Left and right panels correspond to Peter and Carl 
datasets, respectively. Shaded gray areas correspond to standard deviation in the case of the simulated instances. }
\label{ColoracionsPeterSim}	
\end{center}		
\end{figure*}

From our graph collection -see section \ref{Construction}-, we obtain two sequences of chromatic numbers
$s_P(\chi)$ and $s_C(\chi)$ corresponding to the evolution of the chromatic number 
in Peter and Carl datasets,  respectively:
\begin{eqnarray}
s_P(\chi)&=&\chi({\cal G}_{P1}),...,\chi({\cal G}_{P11})\nonumber\\
s_C(\chi)&=&\chi({\cal G}_{C1}),...,\chi({\cal G}_{C11})\nonumber \quad.
\end{eqnarray}
The above sequences display similar patterns with some interesting differences, see 
figure (\ref{ColoracionsPeterSim}) and figure (\ref{Coloracions}).  For example,  the middle stages of both 
datasets  show an increase in the chromatic number.  At the stage when the syntactic 
spur takes place, Peter's dataset $s_P$ displays a sharp transition from a nearly constant, 
low chromatic  number to a high chromatic number, which is fully consistent with
the emergence of complex syntax.  First tree networks in $s_P$ accept $2$-colorings, i.e., they are
bipartite, see figure (\ref{Fig:Colors}). The grammar at this stage mainly 
 generates pairs of complementary words, like:
\begin{eqnarray}
&&\langle {\rm verb,noun}\rangle\;{\rm or}\nonumber\\
&&\langle {\rm adjective ,noun}\rangle\quad.\nonumber
\end{eqnarray}
Typical productions of this stage are, for example,  "car red" or "{\em horsie} run". This pre-transition pattern, also-called $2$-word stage,
corresponds to a highly restrictive grammar,  e.g., syntactic structures like $\langle {\rm verb,verb}\rangle$ 
do not exist.  Instead, relations between lexical items are strongly constrained
by their semantic content.  On the other hand,  Carl's sequence  $s_C$ shows $\chi \geq 3$ 
from the very beginning --i.e. these networks are not bipartite. A detailed inspection of  Carl's productions at this stage shows  the presence of functional particles from the very beginning. This suggests that, in general, high chromatic numbers relates to high grammar flexibility, being this flexibility provided by the {\em hinge} role that have these particles in the global functioning of grammar. 

Still, the behaviour of  $\chi({\cal G})$ can be quite sensitive to the anomalous
 behaviour of small subgraphs.   For example, the transition of $\chi({\cal G}_{2})=2$ to  
 $\chi({\cal G}_{3})=3$, when Peter is about 23 months old, is due a single triangle in 
 a (largely) bipartite network --see figure (\ref{ColoracionsPeterSim}) left. 
A combination of measurements enables us to assess whether the
chromatic number represents the behaviour of a small number of nodes or
it is the natural outcome of global network features. For example, we can compare $\chi({\cal G})$ 
with the lower bound given by the clique number (\ref{clique}) and the upper bound provided by the
 maximal $K$-core connectivity  (\ref{Kcore}). Therefore, each sequence $s_P(\chi),s_C(\chi)$ 
 will be accompanied by two sequences,  namely $\Omega, \kappa$:
\begin{eqnarray}
\Omega_{P,C}&=&\omega({\cal G}_{P1,C1}),...,\omega({\cal G}_{P11,C11})\nonumber\\
\kappa_{P,C}&=&K^*({\cal G}_{P1,C1}),...,K^*({\cal G}_{P11,C11})\nonumber\quad.
\end{eqnarray}
Since every graph can be associated to a measure of $\chi$ relevance --see eq. (\ref{fchi})--, we will have 
a third sequence of $f$'s for every Peter's graph and another associated to every Carl's graph -see table (\ref{TaulaColoring}).  
For example, figure (\ref{ColoracionsPeterSim}) -top-
shows a clear trend towards  increasing maximum clique and maximum $K$-core 
combined with increased relevance --see  table \ref{TaulaColoring}--, which 
indicates that the final chromatic number cannot be longer associated to
any trivial clique.  In any case, the relevance of the chromatic number as a global complexity 
estimator is much more feasible after the transition.  Both Peter and Carl sequences show
that  the chromatic number is often  close to the clique number -see figure \ref{ColoracionsPeterSim} (top). 
Maximum $K^*$-core size is generally more than twice the maximum clique size 
(see figure \ref{ColoracionsPeterSim} (inset)).   Then, the whole network  structure (or a large part of it) 
has enough connectivity to enable the emergence of  a non-trivial $K$-core structure. This is consistent 
with a manual inspection of grammars that generate a great amount of combinatorial complexity, 
i.e., a rich collection of compatibility relations. 

\subsection{Real syntax versus null model}

\label{Subsec:model}

Here, we compare the evolution of the chromatic number in real and simulated networks.
A data-driven, syntax-free model that generates random child's utterances having the same statistics of 
word production as Peter and Carl datasets  is used as a null model \cite{Corominas-Murtra:2009a}.
This model definition enables us to assess if the high combinatorics displayed by 
post-transition networks emerge directly from an increasingly rich vocabulary. We build our model 
by extracting the following  statistical parameters from the $11$ recorded  conversations 
in Peter and Carl corpora: 
\begin{enumerate}
\item
The number of sentences $|S_P(i)|$, $S_C(i)$  in the Peter and Carl datasets.
\item
The probability distribution of {\em  structure lengths} or the probability $P(s)$ 
that any syntactic structure has $s$ words.  We obtain two different distributions,
one for each dataset.
\item
We  assume  that  the probability  of  the $i$-th  most frequent word is a scaling law:
\begin{equation}
  p(i) = {1 \over Z} i^{-\beta}\quad,
\end{equation}
\noindent
with  $1  \le  i  \le  N_w(T)$,  $\beta \approx  1$  --i.e., Zipf's law--
and  $Z$  is  the normalization constant:
\begin{equation}
  Z = \sum_{i=1}^{N_w(T)}\left(\frac{1}{i}\right)^{\beta}\quad.
\end{equation}
Notice that $Z$ depends on lexicon size, $N_w(T)$, which grows slowly at this  stage. 
\end{enumerate}
\begin{figure}
\includegraphics[width=7.5cm]{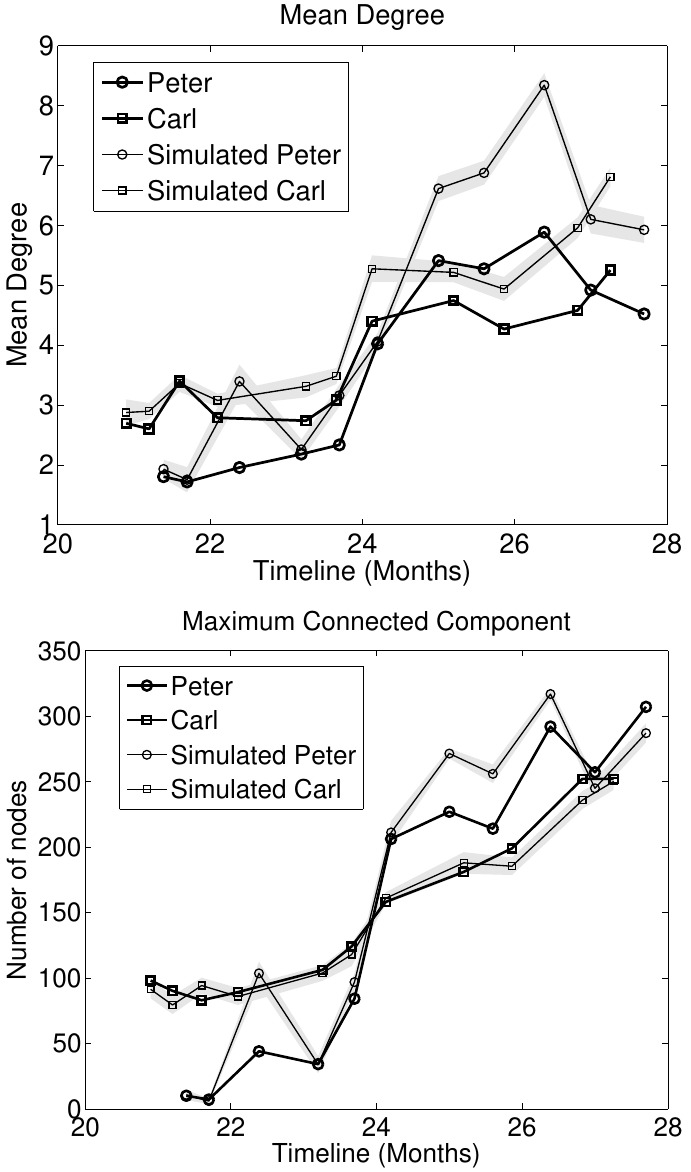}			%
\caption{Evolution of (top) the mean degree and (bottom) size of the largest 
connected component in the real (strong solid lines) and simulated (weak solid lines) syntax networks. Shaded gray areas correspond to standard deviation in the case of the simulated instances. }
\label{GccANDMeanK}	
\end{figure}
\begin{figure*}
\includegraphics[width=12.5cm]{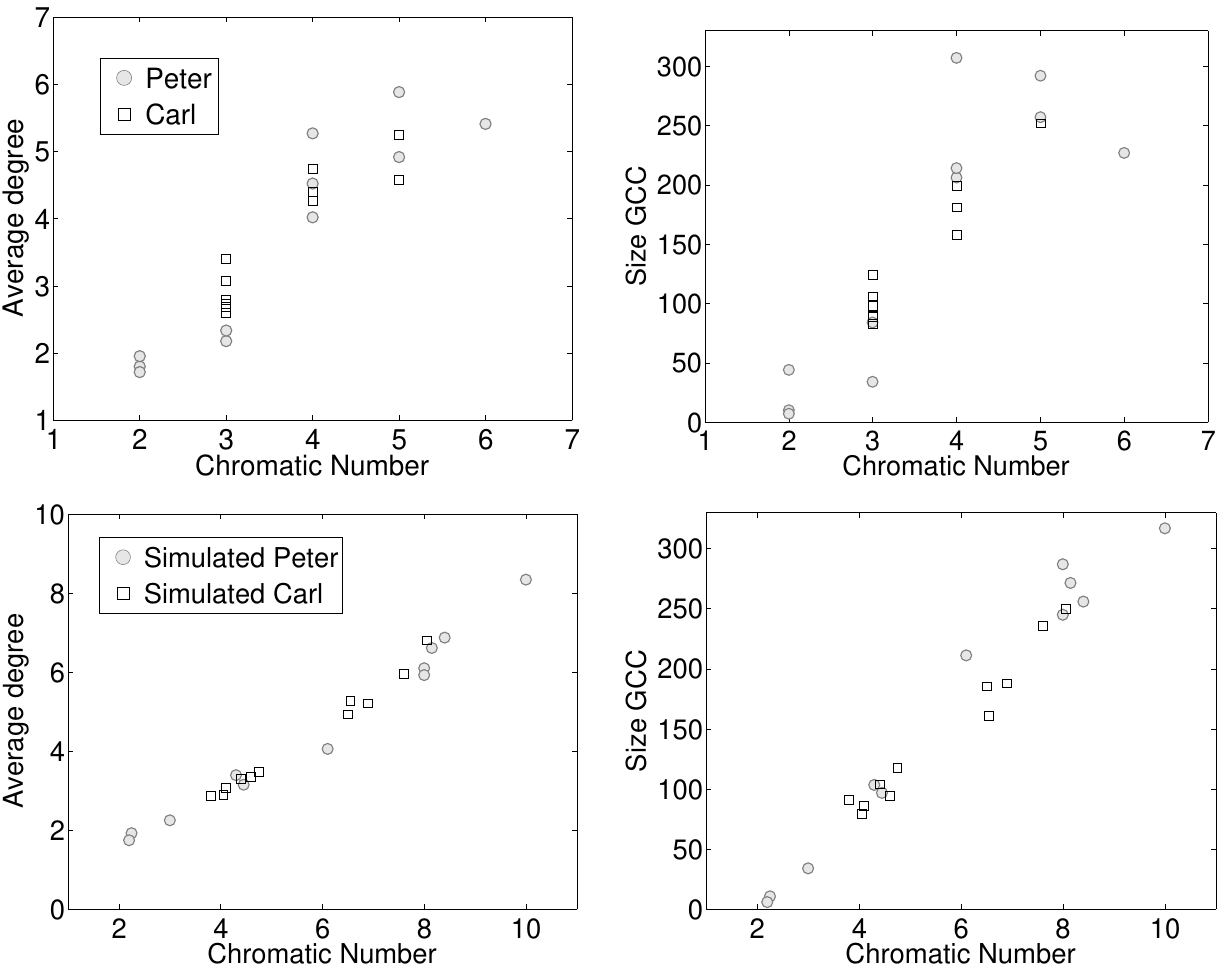}			
 \caption{Relationship between (right) the average degree and the chromatic number and
  (right) the size of the largest connected component and the chromatic number in the Peter
  (solid circles) and Carl (open squares). Simulation results are at the bottom while real
  data is shown at the top.}
\label{Coloracions}	
\end{figure*}

We run the above model in the two datasets by generating $|S_{P,C}(i)|$ random sentences, 
each experiment is repeated $20$ times.  From the collection of randomly generated syntactic structures
we construct a comparable 
sequence of syntax networks following the same method as in the real datasets -see section \ref{Construction}. 
Figure (\ref{GccANDMeanK}) shows that our model generates random syntax networks with size 
and connectivity comparable to the ones measured in real networks.  These statistical indicators 
display a huge increase during the studied period, being this increase sharper around the age
 of two, i.e., during the syntactic spurt \cite{Corominas-Murtra:2009a}. As discussed in section 
II both the mean connectivity and network size play an important role
 when determining the values of $\omega$, $\chi$ and $K^*$. 

Now, we compute the sequence of averaged chromatic numbers, $\tilde{s}_{P}(\chi), 
\tilde{s}_C(\chi)$, for the simulated Peter and Carl syntax networks. Similarly, we 
generate the sequences of average clique number $\tilde{\Omega}_{P,C}$ and 
the average maximum $K$-core $\tilde{\kappa}_{P,C}$. 
The most salient property we find when comparing real networks obtained from both Peter 
and Carl's corpora with their randomized counterparts is a huge increase 
of $\chi, \omega$ and $K^*$ in the simulated networks. That is, the ensemble of random 
strings displays higher complexity parameters than the real corpora. For example, at 
the end of the studied period,  the three complexity estimators are 
close to $10$ in Peter simulations and close to $9$ Carl simulations 
-see figure (\ref{ColoracionsPeterSim})  bottom. 

A very interesting feature is found at the first stages of the simulated Peter sequence: the
random networks are no longer bipartite -see section \ref{Real}. In particular, the third corpus has
an average  chromatic number of $4$, which is significantly higher than the observed
 chromatic number.  In this case, the two-stage grammar imposes severe constraints
on what is actually  plausible in any pre-transition syntactic structure. This trend is 
also observed at latter stages of language acquisition.
In general, simulated networks have higher chromatic numbers than  empirical
networks, although both two types of networks have similar connectivities -by definition.
In some cases, the average chromatic number  of the graphs belonging to the random ensemble is twice the 
real one, see figure (\ref{ColoracionsPeterSim}).  
To better understand the nature of these  deviations, we have compared the behaviour
of chromatic numbers against mean  connectivity and the size of the largest connected
component. Figure \ref{Coloracions} shows a well-defined, non-trivial deviation between 
real networks and random networks.  For example, the plot of chromatic number and network 
size display a quasi linear relationship in simulated networks -see figure (\ref{Coloracions}) 
bottom. These plots suggest that the chromatic number is capturing essential combinatorial  
properties of the underlying system, which  cannot be reproduced with a simple, syntax-free 
random generation model.

\section{Discussion}
\label{Discussion}

Syntax is a characteristic, complex and defining feature of language organization. 
It pervades its capacity for unbounded generative power of the linguistic system \cite{Chomsky:1988}, allows sentences to be organized in 
highly structured ways and is acquired in almost full power by children after being 
exposed to a limited repertoire of examples. Syntax is also one aspect of the whole: 
semantic and phonological aspects need to be taken into account, and they are 
all embedded in (and run by) a cognitive, brain-embodied framework \cite{Jackendoff:2002}. Because of the 
dominant role played by how words actually interact with each other, computational 
and theoretical approaches dealing with word inventories or other statistical trends 
ignoring interactions are likely to be limited. Following previous work that takes advantage 
of complex networks approaches to language organization \cite{Sole:2010} we have 
made a step further in studying the organization of syntax graphs using graph coloring. 
The motivation of this approximation is twofold. On the one hand, graph colorability allows 
to properly detect correlations that are not capture by topological approaches. On the other 
hand, it seems a natural way to substantiate previous claims connecting syntax with 
compatibility relations common with other types of systems, such as chemical structures. 
Since graph coloring naturally defines compatibility through the presence or absence of a common 
label to every pair of nodes, it seems the right framework to study the process of network 
growth in child language. 

The behavior of the 
chromatic number accurately marks the syntactic spurt in language acquisition,
i.e.,  it is a footprint of the generative power of the underlying grammar.
There are limitations associated to the network definition. Syntactic relations are 
structure-dependent, not sequence dependent. Because the network is
an aggregation of text sequences, it cannot fully grasp the hierarchical nature
associated to syntactic constructs.  Still,  the chromatic number is a global
measurement that can detect  grammar constraints by analyzing the pattern of
  network interaction at  different scales. That is, the network representation 
  is an indicator of global linguistic performance and includes some combinatorial
  signal which can be properly detected with the chromatic number. In this
  context, standard network measurements like average degree, clustering or degree distribution are much more
  limited.

There are other, broader implications of our work. The chromatic number can be viewed
as a reciprocal measure of standard community detection. Here, the chromatic number defines
a partition of the network in classes of unlinked nodes. This definition is particularly relevant
in networks where some kind of compatibility relation is at work in the wiring process.
In this case, the standard community structure can be misleading, because elements
of the same class cannot be connected.  The case for syntactic graphs is paradigmatic but the
 partition induced by the chromatic number could shed light into the behaviour of many other systems.
 Additionally, we have proposed to assess the statistical significance of these partitions with the sequence of
  minimal violations -see equation (\ref{fchi}). Future work should
  explore how the chromatic number (and related measures) can be exploited to detect {\em forbidden}
 links in the network. Deviations of the chromatic number (as the ones observed in this paper)
 suggest the presence of combinatorial constraints that must be taken into account, for example,
 when defining proper null-models.


\begin{acknowledgments}
We thank  Complex Systems Lab members for fruitful conversations.
This  work   was  supported by the James McDonnell Foundation (BCM, SV, RSV) and the EU program FP7-ICT- 270212 (MSF).
 \end{acknowledgments}




\end{document}